\documentclass[conference]{ieeeconf}
\IEEEoverridecommandlockouts

\usepackage{cite}
\usepackage{textcomp}

\usepackage{hologo}
\usepackage{listings}

\usepackage{pgfplots}
\usepackage{pgfplotstable}
\def\BibTeX{{\rm B\kern-.05em{\sc i\kern-.025em b}\kern-.08em
    T\kern-.1667em\lower.7ex\hbox{E}\kern-.125emX}}

\usepackage{xcolor}
\usepackage[T1]{fontenc} 
\usepackage[utf8]{inputenc} 
\usepackage[english]{babel} 

\usepackage{amsmath} 
\usepackage{amsfonts}
\usepackage{amssymb}  

\usepackage{diagbox}
\usepackage{makecell}
\usepackage{multirow}

\let\labelindent\relax
\usepackage[inline]{enumitem}

\usepackage{fontawesome}
\usepackage{bclogo}

\usepackage{booktabs} 

\usepackage{dsfont}
\usepackage{graphicx}
\usepackage{subcaption} 

\usepackage{amsmath}
\usepackage{pifont}
\usepackage{booktabs}
\usepackage{siunitx}
\usepackage{tikz}
\usetikzlibrary{shapes.geometric, calc}

\newcommand\score[2]{%
  \pgfmathsetmacro\pgfxa{#1 + 1}%
  \tikzstyle{scorestars}=[star, star points=5, star point ratio=2.25, draw, inner sep=0.15em, anchor=outer point 3]%

  \begin{tikzpicture}[baseline]
    \foreach \i in {1, ..., #2} {
      \pgfmathparse{\i<=#1 ? "yellow" : "gray"}
      \edef\starcolor{\pgfmathresult}
      \draw (\i*1em, 0) node[name=star\i, scorestars, fill=\starcolor]  {};
    }
    \pgfmathparse{#1>int(#1) ? int(#1+1) : 0}
    \let\partstar=\pgfmathresult
    \ifnum\partstar>0
      \pgfmathsetmacro\starpart{#1-(int(#1)}
      \path [clip] ($(star\partstar.outer point 3)!(star\partstar.outer point 2)!(star\partstar.outer point 4)$) rectangle 
      ($(star\partstar.outer point 2 |- star\partstar.outer point 1)!\starpart!(star\partstar.outer point 1 -| star\partstar.outer point 5)$);
      \fill (\partstar*1em, 0) node[scorestars, fill=yellow]  {};
    \fi
  \end{tikzpicture}%
}

\usepackage{tikz}
\usepackage{pgfplots}
\pgfplotsset{compat=newest}
\usepgfplotslibrary{fillbetween}

\usepackage{epsfig} 
\usepackage{times} 
\usepackage{siunitx}
\usepackage{calc}
\usepackage[hyphens]{url}
\usepackage{svg}
\renewcommand{\baselinestretch}{0.90}

\usepackage{fontawesome}

\usepackage{tabularx}
\usepackage{multirow}
\usepackage{pifont}

\usepackage{enumitem}

\usepackage{tabularx}
\usepackage{bbm}
\usepackage{dsfont}

\usepackage{scalerel}
\usepackage{tikz}
\usetikzlibrary{svg.path}

\definecolor{orcidlogocol}{HTML}{A6CE39}
\tikzset{
	orcidlogo/.pic={
		\fill[orcidlogocol] svg{M256,128c0,70.7-57.3,128-128,128C57.3,256,0,198.7,0,128C0,57.3,57.3,0,128,0C198.7,0,256,57.3,256,128z};
		\fill[white] svg{M86.3,186.2H70.9V79.1h15.4v48.4V186.2z}
		svg{M108.9,79.1h41.6c39.6,0,57,28.3,57,53.6c0,27.5-21.5,53.6-56.8,53.6h-41.8V79.1z M124.3,172.4h24.5c34.9,0,42.9-26.5,42.9-39.7c0-21.5-13.7-39.7-43.7-39.7h-23.7V172.4z}
		svg{M88.7,56.8c0,5.5-4.5,10.1-10.1,10.1c-5.6,0-10.1-4.6-10.1-10.1c0-5.6,4.5-10.1,10.1-10.1C84.2,46.7,88.7,51.3,88.7,56.8z};
	}
}

\newcommand\orcidicon[1]{\href{https://orcid.org/#1}{\mbox{\scalerel*{
				\begin{tikzpicture}[yscale=-1,transform shape]
					\pic{orcidlogo};
				\end{tikzpicture}
			}{|}}}}

\usepackage{hyperref} 
\hypersetup{
    colorlinks=true,
    breaklinks=true,
    linkcolor=blue,
    filecolor=magenta,      
    hypertexnames=true,
    urlcolor=cyan,
    pdfpagemode=FullScreen,
    citecolor=green
    }
\usepackage{cleveref}

\title{\LARGE \textbf{A Concept for Efficient Scalability of Automated Driving Allowing for Technical, Legal, Cultural, and Ethical Differences}}

\author{Lars Ullrich$^1$ \orcidicon{0009-0001-8166-3118} , Michael Buchholz$^2$ \orcidicon{0000-0001-5973-0794} , Jonathan Petit$^3$ \orcidicon{0000-0002-8644-1442} , Klaus Dietmayer$^2$ \orcidicon{0000-0002-1651-014X} ,~\IEEEmembership{Senior Member,~IEEE,} \\and Knut Graichen$^1$ \orcidicon{0000-0003-2865-8093} ,~\IEEEmembership{Senior Member,~IEEE}
	\thanks{This research is accomplished within the project ”AUTOtech.agil” (FKZ 01IS22088Y, FKZ 01IS22088W). We acknowledge the financial support for the project by the Federal Ministry of Education and Research of Germany (BMBF).}
	\thanks{$^1$ Chair of Automatic Control, Friedrich-Alexander-Universität Erlangen-Nürnberg (FAU), Cauerstraße 7, 91058 Erlangen, Germany {\tt\footnotesize \{lars.ullrich, knut.graichen\}@fau.de}}
	\thanks{$^2$ Institute of Measurement, Control and Microtechnology, Ulm University, Albert-Einstein-Allee 41, 89081 Ulm, Germany {\tt\footnotesize \{michael.buchholz, klaus.dietmayer\}@uni-ulm.de}}
    \thanks{$^3$ Qualcomm Technologies, Inc., Boxborough, MA, USA {\tt\footnotesize petit@qti.qualcomm.com}}}

\usepackage{hologo}
\usepackage{listings}
\begin{document}

\twocolumn[
\begin{@twocolumnfalse}
	\Huge {IEEE copyright notice} \\ \\
	\large {\copyright\ 2025 IEEE. Personal use of this material is permitted. Permission from IEEE must be obtained for all other uses, in any current or future media, including reprinting/republishing this material for advertising or promotional purposes, creating new collective works, for resale or redistribution to servers or lists, or reuse of any copyrighted component of this work in other works.} \\ \\
	
	{\Large Accepted to be published at the \emph{2025 28th IEEE International Conference on Intelligent Transportation Systems (ITSC)}, Gold Coast, Australia, November 18 – 21, 2025.} \\ \\
	
	Cite as:
	
	\vspace{0.1cm}
	\noindent\fbox{%
		\parbox{\textwidth}{%
			L.~Ullrich, M.~Buchholz, J.~Petit, K.~Dietmayer, and K.~Graichen, "A Concept for Efficient Scalability of Automated Driving Allowing for Technical, Legal, Cultural, and Ethical Differences," in \emph{2025 28th IEEE International Conference on Intelligent Transportation Systems (ITSC)}, to be published. 
		}%
	}
	\vspace{2cm}
	
\end{@twocolumnfalse}
]

\noindent\begin{minipage}{\textwidth}
	
\hologo{BibTeX}:
\footnotesize
\begin{lstlisting}[frame=single]
@inproceedings{ullrich2025scalableAD,
	author={Ullrich, Lars and Buchholz, Michael and Petit, Jonathan and Dietmayer, Klaus and Graichen, Knut},
	booktitle={2025 28th IEEE International Conference on Intelligent Transportation Systems (ITSC)},
	title={A Concept for Efficient Scalability of Automated Driving Allowing for Technical, Legal, Cultural, and Ethical Differences},
	address={Gold Coast, Australia},
	year={2025},
	publisher={IEEE. to be published}
}
\end{lstlisting}
\end{minipage}
\setcounter{page}{0}

\maketitle
\thispagestyle{plain}
\pagestyle{plain}
\bstctlcite{IEEEexample:BSTcontrol}

\begin{abstract}
Efficient scalability of automated driving (AD) is key to reducing costs, enhancing safety, conserving resources, and maximizing impact. However, research focuses on specific vehicles and context, while broad deployment requires scalability across various configurations and environments. Differences in vehicle types, sensors, actuators, but also traffic regulations, legal requirements, cultural dynamics, or even ethical paradigms demand high flexibility of data-driven developed capabilities. In this paper, we address the challenge of scalable adaptation of generic capabilities to desired systems and environments. Our concept follows a two-stage fine-tuning process. In the first stage, fine-tuning to the specific environment takes place through a country-specific reward model that serves as an interface between technological adaptations and socio-political requirements. In the second stage, vehicle-specific transfer learning facilitates system adaptation and governs the validation of design decisions. In sum, our concept offers a data-driven process that integrates both technological and socio-political aspects, enabling effective scalability across technical, legal, cultural, and ethical differences.
\end{abstract}
\section{Introduction}\label{introduction}

From a sustainability perspective, which includes the social, economic, and environmental dimensions \cite{kuhlman2010sustainability}, large-scale implementation of automated driving (AD) is essential to exploit benefits fully \cite{crayton2017autonomous}. This implies that automation is needed across various vehicle configurations and environments. Alongside the complexity of the automation itself, this entails numerous additional challenges.

\begin{figure}
	\centering	
	\includegraphics[width=0.95\linewidth]{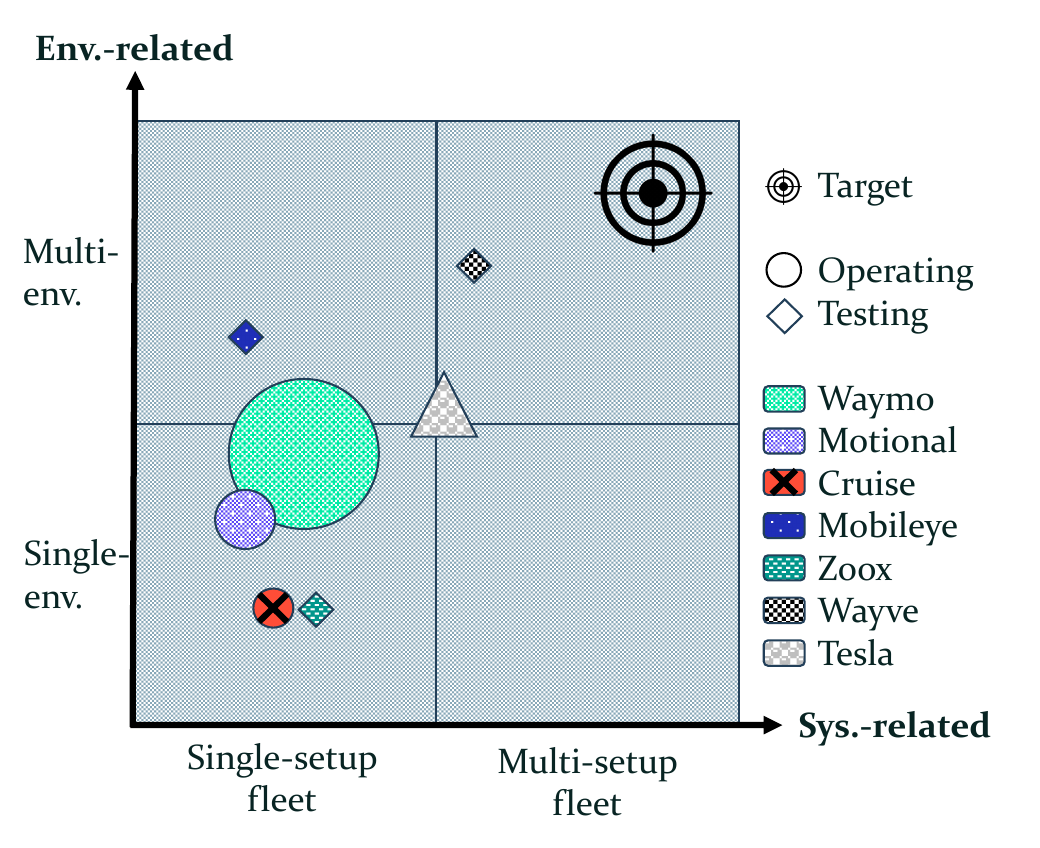}
	\caption{Current state of the art in relation to the scalability goal. Note: Tesla’s Full Self-Driving (FSD) is commercially available but requires constant supervision, placing it between “Operating” and “Testing” (marked with a triangle).}
	\label{fig:current_target}
\end{figure}

At the same time, current research in the field of AD is mainly focused on specific vehicles and environments, as illustrated by \Cref{fig:current_target}. Thereby, artificial intelligence (AI) methods are increasingly being used to enhance the driving capabilities' automation. AI is used both, in sub-modules of modular (M) architectures \cite{carion2020end, scheel2022recurrent}, and end-to-end (E2E) in monolithic architectures \cite{codevilla2018end, jia2023think}. Recently, modular end-to-end (M-E2E) architectures that combine the advantages of the two aforementioned approaches have been proposed \cite{hu2023planning, biswas2024quad}. M-E2E architectures combine differentiated system interpretation, monitoring, and analysis of modularity as well as the mitigation of objective mismatches through an overall E2E alignment. In addition, M-E2E approaches open up a bi-directional, situation-dependent flow of information. This is a key element for replicating human situational awareness \cite{endsley1988design, endsley2020situation} and broader reasoning and thinking \cite{lecun2022path}. 

However, increasing automation through data-driven statistical AI methods requires a paradigm shift towards iterative data-based development as well as verification and validation processes (V\&V) \cite{favaro2023building, ullrich2024expanding} along with data-based safety assurance approaches \cite{diemert2023safety, ullrich2024safetyassurance, ullrich2025datacontrol}. Furthermore, the elevated data dependency limits the direct transferability, and thus, the scalability of the systems. Deviations between the consumed operational data and the characteristics of the development data invalidate assumptions made at development time. Initial work is therefore concerned with the adaptation of sub-tasks \cite{ivanovic2023expanding, ullrich2024transfer} and even the entire task \cite{diehl2024lord} in the event of changes, namely (data) distribution shifts. In addition to the consideration of environmental and system-related distribution shifts, the systematic inclusion of corner cases and unknown unknowns is also crucial. Recognizing corner cases in the real world for different vehicles in different environments and systematically incorporating them into (further) development is decisive for both system credibility and the accounting of temporal changes in the real world. 

Consequently, the key objective for the efficient scalability of AD is the targeted adaptation management of capabilities to system- and environment-related differences. In addition to managing these distribution shifts, the efficient handling of corner cases over the lifecycle is central to addressing both credibility and temporal dynamics in the real world. Moreover, to be not only efficient but also effective, socio-political aspects from trustworthiness to normative and legal conformity must be taken into account in order to be approved by authorities and accepted by society. 

In this paper, we aim to present a concept for efficient scalability that takes into account various differences to enable sustainable AD. Therefore, AD is considered as an extended problem on a global scale. In summary, the contributions of this paper are threefold:
\begin{enumerate}
    \item \textbf{Multi-dimensional problem formulation and analysis:} We present a novel problem formulation for the efficient scalability task of AD.
    \item \textbf{Concept for efficient scalability of AD:} We present a new concept to tackle the problem through a two-stage fine-tuning process, complemented by a iterative data-driven development and refinement.
    \item \textbf{Potential solutions and open questions:}
    We sketch and discuss foreseeable solutions and highlight open questions, outlining future research directions.
\end{enumerate}

\section{Fundamentals}\label{Fundamentals}

\begin{table*}[]
    \centering
    \caption{Comparison of AD stack capabilities across modular (M), service-oriented modular (SO-M), vanilla end-to-end (E2E), modular end-to-end (M-E2E), and service-oriented modular end-to-end (SO-M-E2E) architectures.}
    \resizebox{\textwidth}{!}{
    \label{tab:comparison_AI_method}
    \begin{tabular}{l l l c c c c c c }
        \toprule
        \textbf{AD Stack} & \textbf{Focus} & \textbf{Characteristic} & \textbf{Explainability}$^{1}$ & \textbf{Generalization}$^{2}$ & \textbf{Flexibility}$^{3}$ & \textbf{Adaptation}$^{4}$ & \textbf{Scalability}$^{5}$ \\
        \midrule
        \textbf{M} & task separation &  traceability  & \score{4.5}{5} & \score{1}{5} & \score{2}{5} & \score{3.5}{5} & \score{2}{5}  \\
        
        \textbf{SO-M} & reusable services & agility & \score{4}{5} & \score{2}{5} & \score{4.5}{5} & \score{4.5}{5} & \score{4.5}{5}  \\
        
        \textbf{E2E} & monolithic sense-to-act & AI-driven & \score{0.5}{5} & \score{3.5}{5} & \score{1.5}{5} & \score{2}{5} & \score{2.5}{5}  \\
        
        \textbf{M-E2E} & hybridization & balancing capabilities  & \score{3}{5} & \score{3.5}{5} & \score{2.5}{5} & \score{3}{5} & \score{3.5}{5}  \\
        
        \textbf{SO-M-E2E} & contextual agility & self-orchestration & \score{3.5}{5} & \score{4}{5} & \score{4}{5} & \score{4}{5} & \score{4.5}{5}  \\
        \bottomrule
    \end{tabular}}
    \begin{minipage}[t]{0.49\textwidth}
    \scriptsize
    \begin{itemize}
    \item[$^{1}$] Ability to provide understandable justifications for system behavior.
    \item[$^{3}$] Responsiveness to internal constraints, e.g., processing resources. 
	\end{itemize}
    \end{minipage}
    \hfill
    \begin{minipage}[t]{0.47\textwidth}
    \scriptsize
    \begin{itemize}
    \item[$^{2}$] Ability to perform robust beyond training data, e.g., unseen scenarios. 
    \item[$^{4}$] Capability to adjust to external changes, e.g., varying environments.
	\end{itemize}
    \end{minipage}
    \begin{minipage}{\textwidth}
	\scriptsize
	\begin{itemize}
    \item[$^{5}$] Ability to expand efficiently across different applications, e.g., vehicle types, markets.
	\end{itemize}
	\end{minipage}
    \label{tab:tab1}
\end{table*}

\textbf{AD Stack Evolution.} Traditional hand-crafted architectures are traceable but limited in generalization. While adaptation is possible through modularization, scalability is restricted due to limited generalization and manual adaptation efforts. In contrast, purely AI-driven architectures (E2E) \cite{bojarski2016end, codevilla2019exploring, chen2021learning} can offer improved functional performance and generalization but at the expense of flexibility, adaptability, and explainability. Thus, while E2E is less explainable, it requires extensive retraining to adapt. Therefore, M-E2E architectures that combine the merits are currently popular in research. Unlike this, software-defined vehicles and service-oriented modular (SO-M) architectures \cite{hellmund2016robot, becker2021safety} are aspiring in industry. The flexibility, adaptation, and scalability benefits are crucial for efficiency and economic reasons. 

Consequently, merging service-oriented approaches with M-E2E research represents a sweet spot. Query-based M-E2E approaches, such as \cite{hu2023planning, jiang2023vad, zheng2025genad}, form a starting point as attention-based queries are learnable adaptive interfaces that enable contextual agility, similar to classical services \cite{kampmann2019dynamic}. Furthermore, query-based services are able to incorporate tokenized information, facilitating the integration of large language model (LLM) \cite{cui2023drivellm} and vision language model (VLM) \cite{tian2024drivevlm} capabilities. 

The combination of industry-driven SO-M \cite{hellmund2016robot, becker2021safety} and research-driven M-E2E AD stacks lead to service-oriented modular end-to-end (SO-M-E2E) stacks \cite{ullrich2025adstack}, while easing the linkage of multi-modality, vehicle-to-everything (V2X), and foundation model (FM) research. Apart from the improved capabilities illustrated in \Cref{tab:tab1}, the realization and bridging with areas, such as V2X and FM, is part of future research \cite{you2024v2x, ullrich2025adstack}. As basis for this paper, the latest (SO)-M-E2E approaches, such as \cite{hu2023planning, jiang2023vad, zheng2025genad}, are considered in general terms. This acknowledges that AD stack research has not yet converged, and maintains flexibility through AD stack abstraction to ensure the developed concept remains valid.

\textbf{Paradigm Shifts in Processes.} The increasing automation through data-driven statistical AI urges a variety of paradigm shifts. Transitioning from mathematically explicit to data-based implicit systems affects not only the development and V\&V processes, but also system safety, compliance, release, and operational processes, among others. Thereby, from a methodological, technical, and regulatory perspective, a shift towards iterative product and process refinement is essential. Temporal dynamics and open long-tail distribution of real-world scenarios \cite{liu2019large} amplify this lifecycle and post-market monitoring need. 

\begin{figure}
	\centering	
	\includegraphics[width=0.99\linewidth]{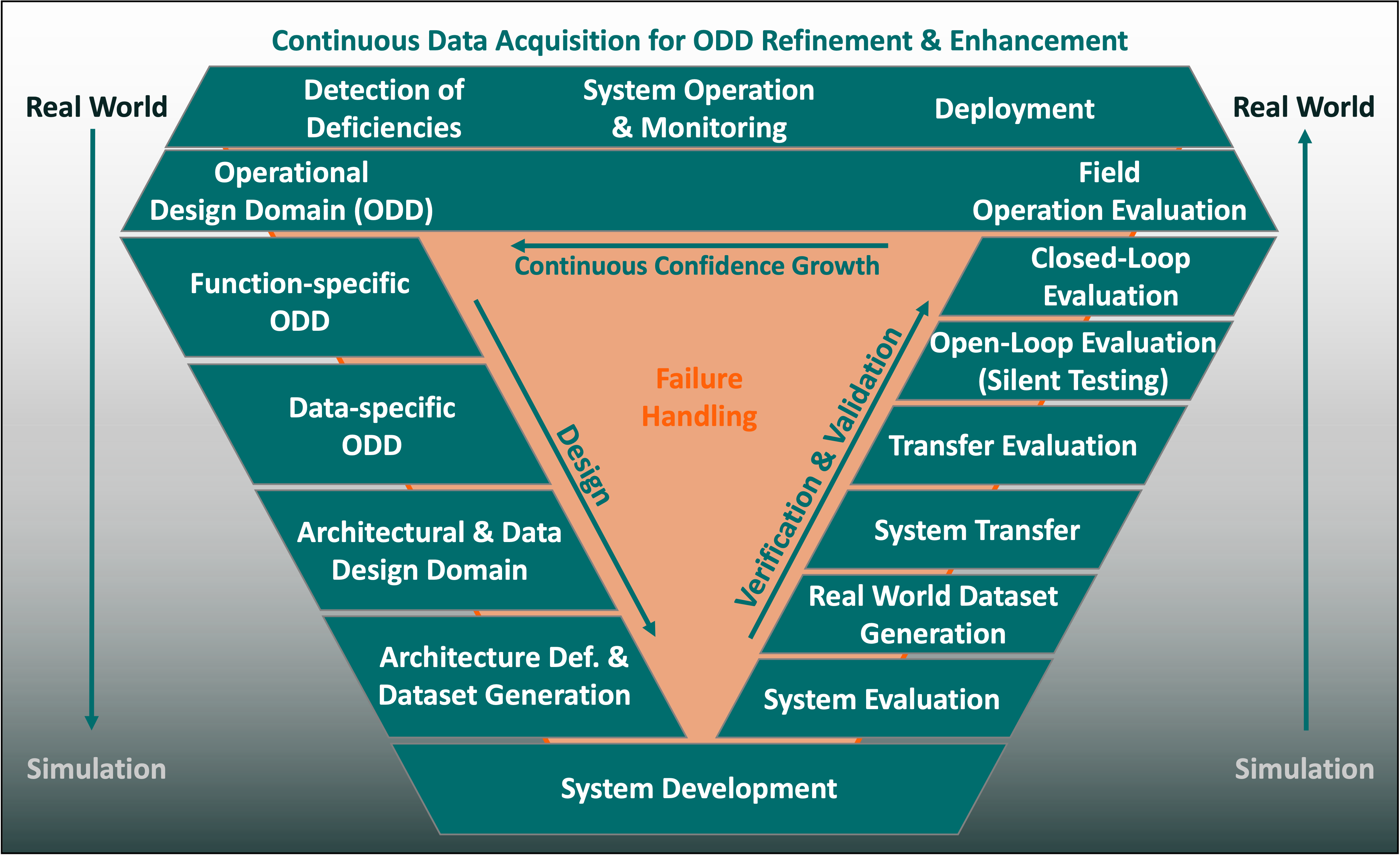}
	\caption{Iterative data-based V-model according to \cite{ullrich2024expanding}, which illustrates data-driven development and V\&V of complex systems incorporating AI.}
	\label{fig:v-model}
\end{figure}

While continuous integration and development (CI/CD) \cite{shahin2017continuous} is proven in classical software engineering, AI engineering also pushes for a shift towards data-based workflows to reflect the underlying characteristic change. As a result, established concepts like Safety of the Intended Functionality (SOTIF) \cite{iso21448} or Safety Integrity Level (SIL) \cite{iso61508} need to be rethought in the context of data-based safety assurance \cite{diemert2023safety, ullrich2024safetyassurance, ullrich2025enhancingtrust, ullrich2025datacontrol}. In addition, due to the empirical nature of AI system development, and especially given safety-critical real world applications, profound simulation-inclusive processes along transitions from and to the real world are key. Automation and AI are also becoming integral to these processes, as demonstrated by Tesla's Data Engine \cite{karpathy_cvpr21}. Consequently, complex systems including AI, such as (SO-)M-E2E-based AD stacks, require adaptive processes. The iterative data-based V-model \cite{ullrich2024expanding}, illustrated in \Cref{fig:v-model}, builds upon the classical V-model \cite{brohl1993v} while harmonizing it with sophisticated processes \cite{favaro2023building, karpathy_cvpr21}. This harmonized process model serves as basis for the system development process of emerging AD systems.

The process outlined in \Cref{fig:v-model} is elaborated in more detail in \cite{ullrich2024expanding}. Without going into detail here, it is crucial to acquire datasets that capture the application in a profound way. In addition to challenges like creating balanced datasets, factors such as temporal dynamics and the open long-tail distribution of real scenarios contribute to the inherent incompleteness of datasets. The early and constant detection of trigger conditions is therefore elementary not only for the first launch but also throughout the product lifetime. In terms of scalability, dealing with data incompatibilities and the coexistence of trigger conditions resulting from driving tasks, vehicle configurations or environmental differences are crucial.
These fundamental data-driven implicit challenges, often overlooked when considering a single system-environment setup, have a significant impact on scalability in both space and time.

\textbf{Addressing Distribution Shifts.} Changes in statistical properties between development and operational data, so-called distribution shifts \cite{storkey2008training}, invalidate assumptions and form the basis for understanding the challenges of data-dependent systems. From a theoretical perspective, various types of distribution shifts (see \Cref{tab:comparison_shifts}) exist. However, the theoretical understanding has limited practical applicability as AI is primarily used for complex tasks. In such cases, the underlying data-generating process \cite{pearl2009causality} is often unknown or unobservable, with latent variables, confounding factors, or Simpson's paradox introducing additional challenges.

While causal learning \cite{gopnik2007causal} offers an approach to address these challenges, statistical AI systems remain dominant. However, these systems are sensitive to distributional shifts, as they typically assume independent and identically distributed (i.i.d.) data. When this assumption is violated, performance can degrade rapidly and radical \cite{bissoto2024even}. Therefore, in practice, it is more important to recognize a shift than to determine its specific nature, with theoretical knowledge often being simplified to the generalized case of domain shift \cite{zhang2013domain}.
\begin{table}[]
	\centering
	\caption{Comparison of distribution shifts between development $P_{dev}$ and operation $P_{ops}$ data distributions with input $X$ and output $Y$.}
    \resizebox{\linewidth}{!}{
		\label{tab:comparison_shifts}
		\begin{tabular}{l l  }
			\toprule
			\textbf{Shift}  & \textbf{Description} \\
			\midrule
            Covariate & $ P_{dev}(X) \neq P_{ops}(X), P_{dev}(Y|X)=P_{ops}(Y|X)$   \\
            Target &  $ P_{dev}(Y) \neq P_{ops}(Y), P_{dev}(X|Y) = P_{ops}(X|Y) $     \\
            Conditional & $P_{dev}(X|Y)\neq P_{ops}(X|Y), P_{dev}(Y) = P_{ops}(Y)$     \\
            Generalized Target & $P_{dev}(X|Y)\neq P_{ops}(X|Y), P_{dev}(Y) \neq P_{ops}(Y)$     \\
            Concept & $P_{dev}(Y|X)\neq P_{ops}(Y|X), P_{dev}(X) = P_{ops}(X)$    \\
            Domain &  $P_{dev}(X,Y)\neq P_{ops}(X,Y)$    \\
			\bottomrule \\[-6pt]
	\end{tabular}}
\end{table}

Depending on the task, different methods are recommended to handle distribution shifts, as shown in \Cref{tab:comparison_shifts_counters}, with each method offering distinct pros and cons. For instance, domain adaptation \cite{ganin2015unsupervised} and transfer learning \cite{ zhuang2020comprehensive} require training multiple models for different domains, while domain generalization \cite{zhou2022domain} aims to maintain performance across domain shifts. However, achieving generalization across all shifts is difficult, and in some cases, different system behaviors may be preferred, conflicting with the generalization principle.

\begin{table}[]
	\centering
	\caption{Comparison of AI categories with regard to source/target ($\mathcal{S/T}$) shifts on joint distribution ($P_{XY}$) and label space ($Y$) as well as general settings like the number of source domains/tasks ($K$) and the availability of target marginal ($P_{X}^{\mathcal{T}}$) based on \cite{zhou2022domain}.}
	\resizebox{\linewidth}{!}{
		\label{tab:comparison_shifts_counters}
		\begin{tabular}{p{3.5cm}  c c c  c c c}
			\toprule
			\textbf{Architecture}  & \multicolumn{3}{c}{\textbf{Settings}} & \multicolumn{3}{c}{\textbf{Shifts}} \\
			&& K & $P_{X}^{\mathcal{T}}$ &&  $Y_{S/T}$  & $P_{XY}^{\mathcal{S/T}}$   \\
			\midrule
			Supervised Learning && $=1$ & N/A  && $=$ & $=$  \\
			Multi-Task Learning && $> 1$ & N/A && $=$& $=$   \\
			Transfer Learning && $\geq 1$ & Avail. && $\neq$ & $\neq$  \\
			Zero-Shot Learning && $= 1$ & N/A && $\neq$ & $\neq$ \\
			Domain Adaptation && $\geq 1$ & Avail.&& $=, \neq$ &  $\neq$ \\
			Test-Time Training && $\geq 1$ & Partial &&  $=$  & $\neq$ \\
			Domain Generalization && $\geq 1$ & N/A && $=, \neq$ & $\neq$  \\
			\bottomrule \\[-6pt]
	\end{tabular}}
    \begin{minipage}{0.99\columnwidth}
    \centering
    \scriptsize
    \textit{Avail.=Available, N/A-=Not Available, Partial=Partial Available.}
    \end{minipage}
\end{table}

\textbf{Summary.} In AD, distribution shifts arise from incomplete scenario coverage in training datasets and the open long-tail distribution of the real world, which are typically addressed by generalization. However, changes related to scalability, outlined in \Cref{tab:categories}, are difficult to be addressed through generalization. Initial work is therefore concerned with the adaptation of sub-tasks \cite{ivanovic2023expanding, ullrich2024transfer} and even the entire task \cite{diehl2024lord}. Here, current research primarily addresses adaptation as a decoupled problem, thus simplifying the task and overlooking factors outlined in \Cref{tab:categories}. In addition, as the current research builds on existing work, the AD stack evolution and process paradigm shift are neglected. Therefore, this paper aims at structuring the broader multi-dimensional problem while proposing an initial concept for effective and efficient scalability of AD.

\begin{table*}[h!]
\centering
\caption{Overview of changes related to scalability across vehicle setups and environments.}
\resizebox{\textwidth}{!}{
\footnotesize
\renewcommand{\arraystretch}{0.8} 
\setlength{\baselineskip}{0.85\baselineskip} 
\begin{tabularx}{\linewidth}{p{0.5cm} p{1cm} p{3.9cm} p{5.4cm} p{4.7cm}}
\toprule \textbf{Origin} &
\textbf{Category} & \textbf{Aspects} & \textbf{Descriptions} & \textbf{Example}\\ \midrule

\multirow{2}{*}{\rotatebox{90}{\parbox{0.7cm}{\centering Sys.-related}}} & Technical  & \ding{226} Vehicle body & dynamic class, technical constraints & \multirow{2}{=}{Price segment: Hardware and system capabilities vary between vehicles.}
\\ \cmidrule(lr){3-4}
& & \ding{226} Vehicle segment& sensor actuator setup, comp. resources & \\
\midrule 

\multirow{4}{*}{\rotatebox{90}{\parbox{4cm}{\centering Environment-\\ related}}} & Technical  & \ding{226} Traffic infrastructure 
& GPS, internet avail., V2X, traffic control system & \multirow{2}{=}{Road side units: Avail. \& message format (DENM,CAM) varies across regions.} 
\\ \cmidrule(lr){3-4}
& & \ding{226} Market-specific vehicle req.  & 
regulatory compliance, customer preferences \vspace{1mm} & \\
\cmidrule(lr){2-5}

& Legal  & \ding{226} Regulatory frameworks &  AI regulations, automotive regulations & \multirow{2}{=}{AI regulation: Approaches vary across regions and evolve over time.}\\  \cmidrule(lr){3-4}
& & \ding{226} Obligations \& requirements & certification, approval, liability, insurance & \\
\cmidrule(lr){2-5}

& Cultural & \ding{226} Social norms \& cultural expect. &  pedestrian priority, traffic violation tolerance, emergency clearance  & \multirow{2}{=}{Pedestrian interaction: Pedestrian prediction \& interaction differs around the globe.}
\\ \cmidrule(lr){3-4}
& & \ding{226} Cultural behavior &interaction dominance, pedestrian habits, honking norms & \\
\cmidrule(lr){2-5}

& Ethical  & \ding{226} Moral principles \& norm. ethics & consequentialism, deontology, virtue ethics & \multirow{2}{=}{Trolley problem: Moral decision-making in emergencies differs around the globe.} 
\\ \cmidrule(lr){3-4}
& & \ding{226} Ethical standards \& norms & data privacy, human oversight & \\
\bottomrule
\end{tabularx}
\label{tab:categories}}
\end{table*}
\section{Problem Formulation \& Analysis}\label{ProblemFormulation}

\textbf{Target.} The aim is to achieve efficient scalability of AD across different environments and vehicle configurations, taking into account modern (SO)-M-E2E AD stacks and respective processes, while allowing for technical, legal, cultural, and ethical differences.

\begin{figure}[]
	\centering	
	\includegraphics[width=0.995\linewidth]{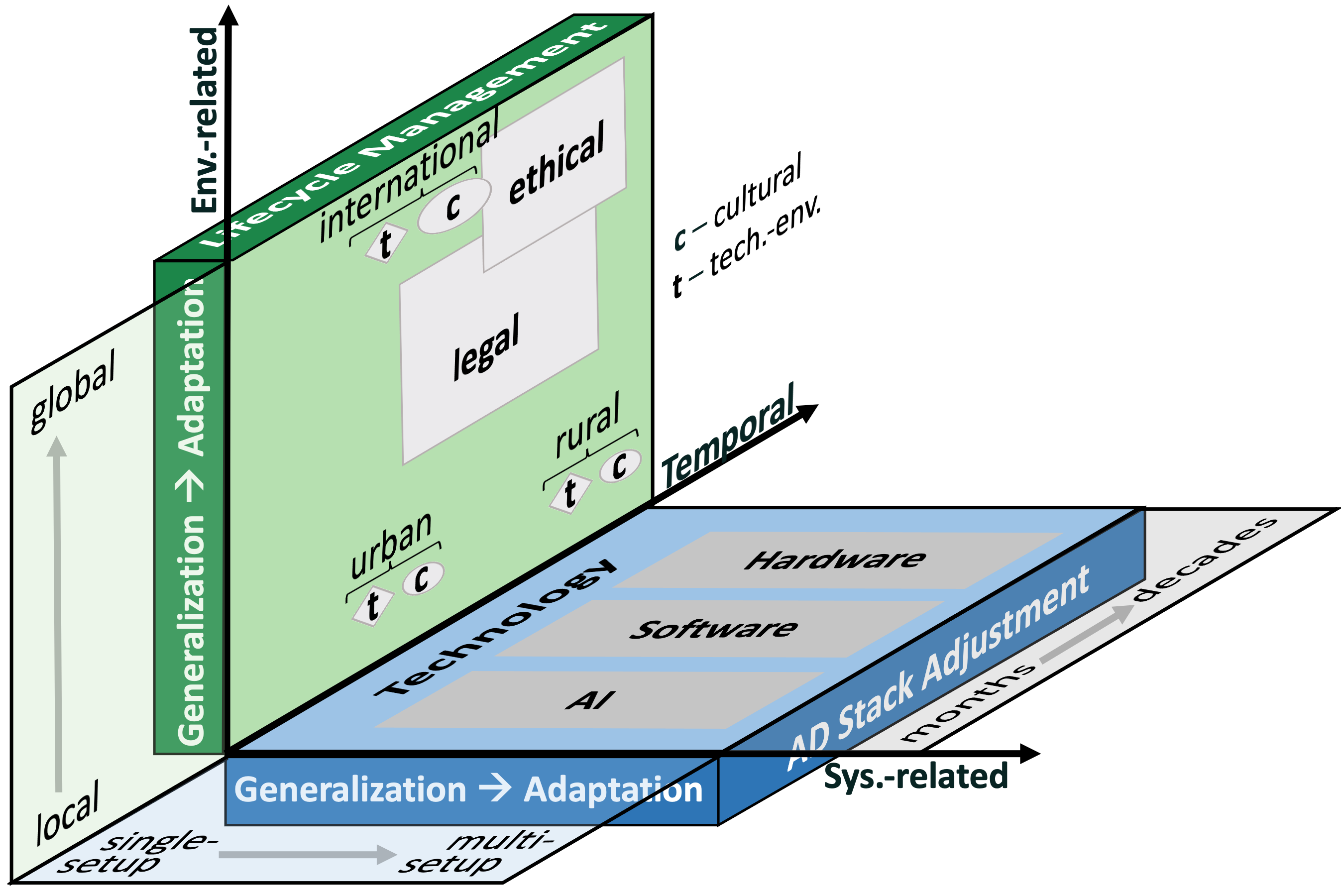}
	\caption{Illustration of the multi-dimensional problem space.}
	\label{fig:multi_prob}
\end{figure}

\textbf{Multi-dimensional Problem Space.} The general problem space is defined by the dimensions of system, space, and time. To avoid unnecessary complexity and derive a targeted solution, the problem space focuses on the key elements: environmental-, and system-related changes along temporal effects. \Cref{fig:multi_prob} illustrates this by projecting the previously discussed key aspects (\Cref{tab:categories}) onto the two primary planes, emphasizing the most significant factors while excluding secondary influences. Despite the existence of decoupled, challenge-specific countermeasures (e.g., generalization, adaptation, AD stack adjsutment, lifecycle management) depicted along the edges of these planes, efficient scalability requires a holistic approach that integrates the multi-dimensional requirements.

Correspondingly, \Cref{tab:categories_countermeasure} provides an overview of the impacts and countermeasures associated with scalability-related changes to the AD system, stack, and processes. The categories of differences are interconnected. Moreover, \Cref{fig:multi_prob} and \Cref{tab:categories_countermeasure} illustrate that an iterative, data-driven process can generically address temporal changes. However, at the same time, the question arises of where to draw the line between generalization and adaptation along both environmental and system-related axes. The logical environmental boundary is given by country borders, while for vehicle configuration, the decisive factor is the sensor setup.

However, the challenge extends beyond adapting along a single axis, it requires a holistic system perspective. Collaborative and collective approaches are crucial, not only to address the inherent incompleteness of datasets but also for prospective and retrospective system and safety analysis.

\begin{table*}[h!]
\centering
\caption{Overview of the effects and countermeasures for scalability-related changes presented in \Cref{tab:categories}.}
\resizebox{\textwidth}{!}{
\footnotesize
\renewcommand{\arraystretch}{0.8} 
\setlength{\baselineskip}{0.85\baselineskip} 
\begin{tabularx}{\linewidth}{p{0.5cm} p{1cm} p{3.3cm} p{5.2cm} p{5.6cm}}
\toprule \textbf{Origin} &
\textbf{Category} & \textbf{Effect on \dots } & \textbf{Addressed through \dots } & \textbf{Example}\\ \midrule

\multirow{2}{*}{\rotatebox{90}{\parbox{0.65cm}{\centering Sys.-related}}} & Technical  & hardware-based capabilities/constraints. & adjusting sensor inputs \& action constraints within the AD stack model.
 & Price segment: Re-train and V\&V for capabilities/constraints at hand. \\ 
\midrule

\multirow{4}{*}{\rotatebox{90}{\parbox{3cm}{\centering Environment-\\ related}}} & Technical  & external information availability, environmental constrains. & dynamic orchestration of various services. & Road side units: DENM- and CAM-based services along orchestration for existence and absence of messages.\\
\cmidrule(lr){2-5}

& Legal  & processes, particularly from testing to operation.  & interchangeable regulatory constraints along the processes for legal conformity.  & AI regulation: Different requirements and obligations across regions.\\  
\cmidrule(lr){2-5}

& Cultural & AD stack sub-modules \& overall task alignment. & exchangeable AD stack sub-modules \& overall fine-tuning for normative compliance. & Pedestrian interaction: Interchangeable prediction modules. Region-specific behavior adaptation.
\\ 
\cmidrule(lr){2-5}

& Ethical  & decision-making logic.  & aligning decision-making to normative ethics. & Trolley problem: Adjustment risk assessment, trade-offs, objectives, and costs.
\\ 
\bottomrule
\end{tabularx}}
\label{tab:categories_countermeasure}
\end{table*}

\textbf{Key Question.} What enables a structured transition of AD systems from a single-setup, single-environment (SSSE) task to a multi-setup, multi-environment (MSME) task across development, deployment, monitoring, and refinement?

\textbf{Observations.}
A full generalization across all configuration setups and environments is neither expected nor appropriate, as certain differences (see \Cref{tab:categories_countermeasure}) necessitate adaptations, e.g., in terms of behavior. At the same time, developing and training a separate model for each environment and vehicle configuration from scratch is not efficiently scalable. Despite multi-dimensional shifts across vehicles, environments, and time, a common foundation remains—progressing safely, normatively compliant, efficiently, and comfortably towards a defined goal.

Furthermore, current (SO)-M-E2E AD stacks primarily focus on core driving functionality, often overlooking external context integration, such as V2X communication. While differences in external information may initially be disregarded, they present valuable opportunities as service-oriented SO-M-E2E AD stacks evolve \cite{ullrich2025adstack}.

Additionally, modularization within these stacks enables rapid adaptation of task-specific skills, e.g., via country-specific pre-trained pedestrian prediction sub-modules. This modularity not only reduces capability adjustment efforts but also facilitates seamless integration of advancements. Recent perception progress, driven by PointPillars \cite{lang2019pointpillars} or DETR \cite{zhu2020deformable} underline the need for continuous backbone and bottom-up updates.

Conversely, top-down E2E alignment \cite{codevilla2019exploring} plays a critical role in eliminating mismatches \cite{eysenbach2022mismatched} while providing a structured framework for incorporating cultural and ethical aspects. In doing so, the task spans from perception to trajectory planning (P2T), which is why we refer to the respective AD stack model as P2T.

\noindent\quad\textbf{Assumptions.} 
\begin{itemize}
    \item If a (SO)-M-E2E model cannot perform well in a generalized simulation environment, it will not perform well in a specific real world setting. 
    \item If a (SO)-M-E2E model cannot perform well under a full set of high-fidelity sensors and highly dynamic actuators (high-end price segment setup), it will not perform well in capability constrainted vehicle configurations.   
    \item[\ding{43}] A pre-trained (SO)-M-E2E P2T base model that performs well in a generic simulation environment under a full set of high-fidelity, multi-modal sensor inputs is presumed, representing generic learned driving capabilities.
\end{itemize}
\noindent\quad\textbf{Guiding Questions.} 
\begin{enumerate}[label=\textbf{Q\arabic*},  left=-0.0em]
  \item \label{q:1}How to align the generic capabilities of the P2T base model to different environments/countries? 
  \item \label{q:2}How to deploy the generic capabilities of the P2T base model on divers vehicle configurations?
  \item \label{q:3}How to integrate trigger conditions detected across environments/countries and system setups in the (further) development.
\end{enumerate}

\section{Concept Development}\label{ConceptDev}

\begin{figure*}
	\centering
    \begin{tikzpicture}
        \node[anchor=south west, inner sep=0] (image) at (0,0) 
            {\includegraphics[width=0.94\linewidth]{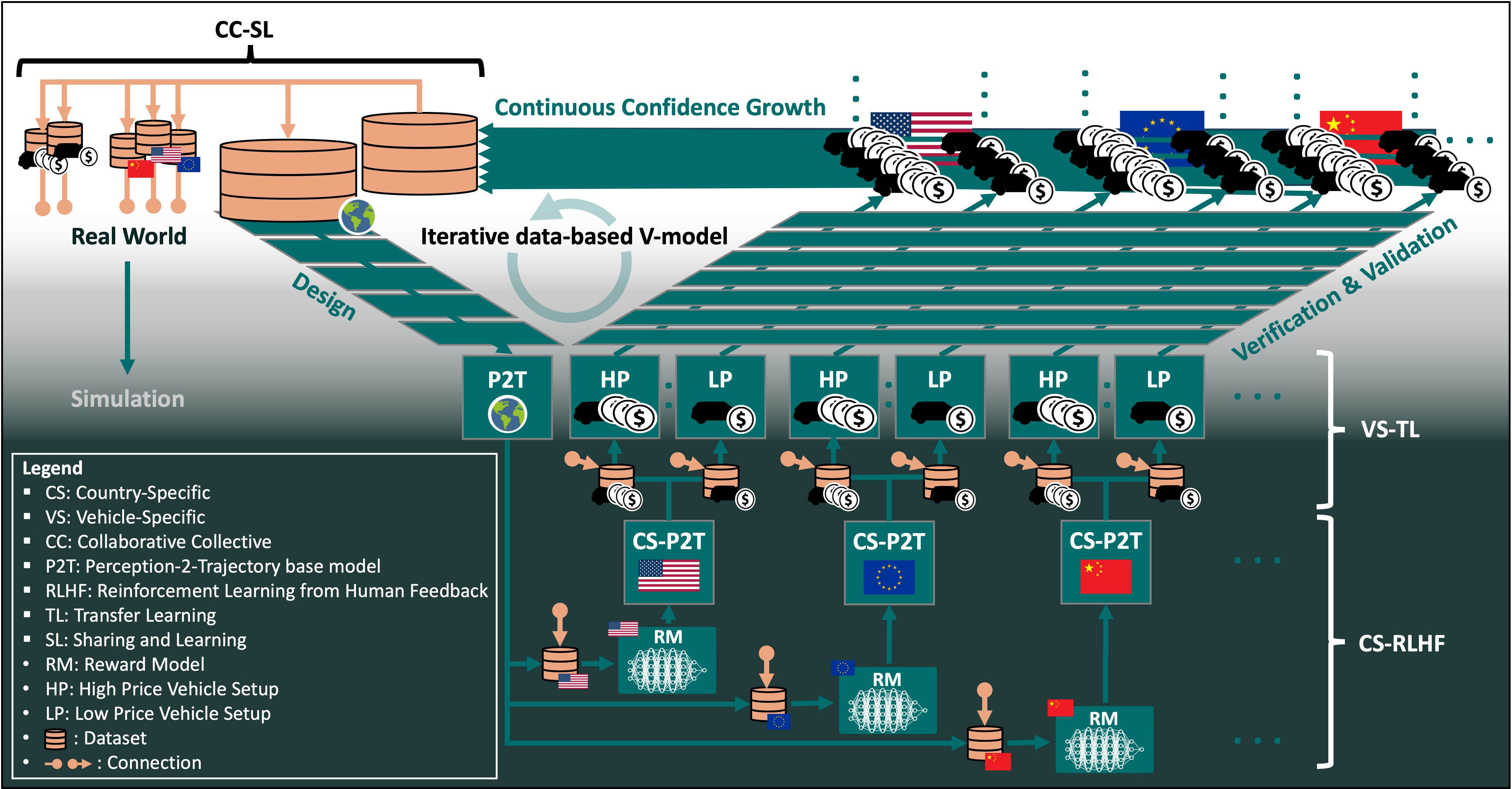}};
        \small
        \node[anchor=north west] at (5.75,5.93){[\Cref{fig:v-model}]};
        \end{tikzpicture}
	\caption{Visualization of the developed concept including the two-stage fine-tuning process.}
	\label{fig:concept}
    \vspace{-1mm}
\end{figure*}

To address the problem outlined in the previous section, we propose transferring established solution strategies from functional-driven SSSE task to the broader scalability-driven MSME task. Thereby, we build upon the assumed pre-trained P2T base model, that has acquired general capabilities in a generic simulation environment.

\textbf{Modularization.} Transferring the modularization strategy is promising, as it enables reusability—key for efficient scalability—while also guiding problem and complexity decomposition. This decomposition should account for both environmental and system-related variations, which drive the need for adaptation.

\textbf{Iterative Refinement.} While closing loops between simulation and real world at the SSSE task level is integral for iterative refinement, MSME requires closing loops between a generic P2T base model and the multitude of specific P2T models under deployment. Otherwise, the MSME task fragments into numerous SSSE tasks, which is contrary to resuability and efficient scalability. 

\textbf{Overall (E2E) Alignment.} AD stack research highlights the necessity of aligning the system with the overall task for success. The broader MSEM task encompasses not only legal, cultural, and ethical driving constraints but also broader socio-political factors. Accordingly, the overall alignment ranges, e.g., from legal frameworks such as the EU AI Act \cite{eu_parliament_2024corr} across authority and societal acceptance up to general trustworthiness.

\textbf{Service-Orientation (SO).} While SO is well established within SSSE-related AD task stacks \cite{hellmund2016robot, kampmann2019dynamic, becker2021safety}, the benefit of flexible adaptation is also key at the broader MSME task level. SO empowers efficient scalability w.r.t. functional-driven concerns, e.g., by switching models when crossing borders, as well as process-driven concerns, e.g., by interchangeable sets of socio-political requirements.

\textbf{Core Concept--Big Picture.}
The combination of the aforementioned solution strategies is depicted within the overall closed loop of development and refinement at the MSME task level in \Cref{fig:concept}. The core is a hierarchical two-stage fine-tuning process along a successive constraint tightening. First, the generic P2T base model is refined to multiple country-specific P2T base models. Second, the country-specific P2T base models are adapted to different vehicle setups before the step-wise transfer to the real system. Besides closing the loop at the general level, country- and setup-specific feedback loops are also included. The sequential decomposition of environmental and vehicle-specific aspects is aligned with the spatial roll-out-first strategy of companies such as Waymo. While within \Cref{fig:concept} the questions (\Cref{q:1}, \Cref{q:2}, \Cref{q:3}) are considered implicitly, the applied principles and respective potential solutions are described in more detail in the following.

Overall, two key aspects should be considered. First, country-specific models could account for a set of countries with strong similarities, such as those within the EU. Second, the order of the fine-tuning process can be reversed, e.g., in case of a single-vendor implementation. For clarity, the fine-tuning order will subsequently be considered as illustrated in \Cref{fig:concept}.

\noindent\quad\textbf{Applied Principles \& Potential Solutions.} 
\begin{itemize}
    \item Like Sudoku \cite{yato2003complexity}, checking a solution--valid trajectory--is less complex (\textsf{P}) than finding a solution (\textsf{NP-C}).
    \item The entangled solution space naively integrates legal, cultural, and ethical aspects while allowing for comprehensive and interpretable assessments.
    \item[\ding{43}]  A top-down approach to fine-tune the base model on top of the solution space facilitates the native consideration of intertwined environmental multiobject constraints. $\Rightarrow$ 
    \begin{enumerate*}[label=\textbf{S\arabic*}]
    \item \label{s:1} \textit{Country-specific reinforcement learning from human feedback (CS-RLHF).}     
    \end{enumerate*}
\end{itemize}

\begin{itemize}
    \item Transitioning from a simulation-based setup to a real world setup reveals trade-offs and establishes requirements to maintain performance.
    \item Reducing sensor modalities and quality affects overall system capabilities, potentially leading to performance degradation or requirement violations.
    \item[\ding{43}] Transfer learning a generic model, developed under an ideal sensor setup, to a constrained real world setup natively reveals hardware requirements while adapting the generic system to the specific application. $\Rightarrow$ 
    \begin{enumerate*}[label=\textbf{S\arabic*}]
    \setcounter{enumi}{1}
    \item \label{s:2} \textit{Vehicle-specific transfer learning (VS-TL).} 
    \end{enumerate*}
\end{itemize}

\vspace{1mm}
\begin{itemize}
    \item If a generic safety-critical corner case is detected, it should be considered across all countries and vehicle setups, independent of detection constraints such as country- or modality-specific aspects.
    \item Crucial setup-specific trigger conditions should be shared across the respective setup-specific community.
    \item Transitioning from databases, such as Fatal Analysis Reporting Systems (FARS), Crash Report Sampling Systems (CRSS), or Crash Investigation Sampling Systems (CISS), to national and global datasets enable native integration into data-driven iterative development processes.
    \item[\ding{43}] Transitioning to streamlined datasets of relevant aspects enables automating closing the loop, thus continuous system and process refinement. $\Rightarrow$  \begin{enumerate*}[label=\textbf{S\arabic*}]
    \setcounter{enumi}{2}
    \item \label{s:3} \textit{Collaborative Collective Sharing and Learning (CC-SL).}
    \end{enumerate*}
\end{itemize}

\section{Concept Core Components}\label{CoreComponents}
The developed core concept is based on three potential solutions (\Cref{s:1}, \Cref{s:2}, \Cref{s:3}), the core components, which are described in more detail in the following.

\textbf{Country-Specific Reinforcement Learning from Human Feedback.} The pre-trained P2T base model, which is already able to generate trajectories, should be post-trained to consider normative compliance and ethics along social norms and cultural expectations among others. Within FM research, such as LLM development, reinforcement learning from human feedback (RLHF) is widely used \cite{christiano2017deep, bai2022training, openai2023gpt4}. Within the given MSME task, RLHF is promising, as it enables to operate on the corresponding meta level, taking into account broad requirements.

In more detail, the core idea is that human experts assign a scalar rating between zero and one to each trajectory across various scenarios, reflecting its desirability. This feedback is then used to train a reward model that mimics human assessments, forming the foundation for P2T base model post-alignment. In order to consider the environmental-related variations, creating one reward model per country, enables to fine-tune the generic P2T base model towards country-specific P2T base models. Benefits of such a CS-RLHF along open research questions are discussed in the following.

\noindent\quad\underline{Benefits \& Opportunities:}
\begin{itemize}
    \item[\faLightbulbO] Eliminate the manual extraction of complex intertwined legal, cultural, or ethical requirements and expectations as well as corresponding cost function design.
    \item[\faLightbulbO] Implicitly provides flexible cost functions, that are dependent on the respective situation and context.
    \item[\faLightbulbO] Tripod of scenarios, trajectories, and human expert ratings facilitate a non-expert comprehensible foundation empowering public accessible safety argumentation, legally required human oversight, socio-political concerns inclusion, along confidence growth.
    \item[\faLightbulbO] Decouples technological dynamics (AI, AD stack) from cultural dynamics (regulation, expectation).
    \item[\faLightbulbO] Enables a streamlined assessment of actions across all ethical levels, from situational and applied ethics to normative and meta-ethics. This ensures a broader discussion of acceptable risk while avoiding semantic gaps.
\end{itemize}

\noindent\quad\underline{Open Research Questions:}
\begin{itemize}
    \item[\faSearch] How to visualize trajectories for effective and efficient rating by experts?
    \item[\faSearch] How many scenarios and trajectories are needed to reach a valid reward model for a country?
    \item[\faSearch] What are the costs of developing and maintaining a country-specific reward model, and who is responsible?
\end{itemize}

Overall, the need for country-specific fine-tuning becomes apparent, as defensive, reactive driving behavior may work well in a cautious environment, but causes a significant performance drop in an aggressive, dominant environment and vice versa. Furthermore, society-inclusive fine-tuning broadens the scope and addresses predominantly unanswered questions related to acceptable risk and critical decisions, such as the trolley problem. Ultimately, CS-RLHF resembles top-down alignment approaches like behavioral cloning (BC) \cite{codevilla2019exploring}, conditional imitation learning (CIL) \cite{codevilla2018end}, and teacher-student models \cite{chen2021learning}. It can be viewed as a reinforcement learning-based teacher-student framework at the meta level, complementing the overall multi-level modeling and alignment.

\textbf{Vehicle-Specific Transfer Learning.} After adapting a generic P2T base model to multiple country-specific versions, refining the models for vehicle-specific characteristics takes effect. Therefore, the ideal sensor setup in the simulation is replaced by the desired vehicle configuration. In addition to sensor modality and quality adjustments, vehicle dynamics are aligned to the real system. Continuous training, aka fine-tuning, of the model under the changed circumstances on overall metrics allows vehicle-specific refinement. Following this initial simulation-based pre-validation of the desired system setup, the transfer process, outlined in \cite{ullrich2024expanding}, proceeds from open-loop (silent testing) to closed-loop, followed by field operations and, ultimately, deployment in the real vehicle.

\noindent\quad\underline{Benefits \& Opportunities:}
\begin{itemize}
    \item[\faLightbulbO] Enables validation of design decisions early on in simulation.
    \item[\faLightbulbO] Facilitates uncovering minimum requirements, trade-offs, and sensitivity of quality levels and/or sensor modalities.
\end{itemize}

\noindent\quad\underline{Open Research Questions:}
\begin{itemize}
    \item[\faSearch] Is it possible to transfer learn capabilities to highly limited sensor setups, such as vision-only systems?
    \item[\faSearch] Is it possible to transfer learn capabilities across major setup changes, e.g., adaptation to heavy-duty systems?
\end{itemize}

It should be noted that, with the exception of Tesla, most industries use a homogeneous sensor setup \cite{ullrich2025adstack}. Moreover, given the cost-intensive nature of AD development and the large market potential, a spatial roll-out-first strategy is often considered. As a result, the relaxation of requirements and the transition to multiple, e.g., lower-cost vehicle setups follow later on.

\textbf{Collaborative Collective Sharing and Learning.} Given a variety of vehicle setups in operation across different countries, collective and collaborative data acquisition becomes possible. In data-driven development, V\&V, approval, and more, the integration of acquired data is crucial for effective refinement. To leverage collected data, disentangling trigger conditions and categorizing them as “general”, “environmental” or “setup” is the first step. The second step is to transfer the data into a desired standardized representation. For instance, in the "general" case, this involves transferring sensor recordings from specific setups and environments into a generic full-sensor simulation environment. The third step involves updating development and V\&V datasets along the two-stage fine-tuning process. With the updated datasets, collective learning can enhance the general PT2 base model, while collaborative learning refines environment- and setup-specific models.

\noindent\quad\underline{Benefits \& Opportunities:}
\begin{itemize}
    \item[\faLightbulbO] Unifying existing databases, e.g., CISS, with relevant ops. data implicitly updates the underlying assumptions.
    \item[\faLightbulbO] Addressing data-driven limitations, such as incompleteness, temporal dynamics, while ensuring hazardous Unknown Unknowns are encountered only once. 
    \item[\faLightbulbO] Providing a foundation for a streamlined and evidence-based retrospective system and safety analysis.
\end{itemize}

\noindent\quad\underline{Open Research Questions:}
\begin{itemize}
    \item[\faSearch] How to disentangle trigger causes in an automatable way? 
    \item[\faSearch] Can Generative Adversarial Networks (GAN) or Generative AI's (GenAI) automate the transition from real world data to simulation-based scenarios and comprehensive synthetic sensor readings?
    \item[\faSearch] Can CC-SL be achieved through open innovation and close collaboration without strict regulatory sharing requirements, while also preventing monopolies?
\end{itemize}

Initial initiatives, such as NXT GEN AI METHODS \cite{NextAIM} and IAV Mela \cite{IAV_Mela} or GAIA-2 \cite{russell2025gaia}, target specific open questions. Additionally, \cite{wang2024does} introduces the concept of imaginative intelligence, which is relevant in this context alongside GANs and GenAI. While some sub-questions are already under discussion, the CC-SL as a whole has not yet been addressed, particularly the promising integration of long-standing databases.

Overall, CC-SL serves as a key enabler of efficient scalability while relying on the sharing of data-based assets. The level of sharing can vary based on agreements and partnerships, with extensive exploitation requiring an intelligent knowledge-sharing system. Federated learning, for example, enables knowledge exchange without fully disclosing data. Notably, sharing is mutually beneficial, as it provides access to a broader and more comprehensive dataset. Additionally, integrating official databases raises further questions about data ownership and control.

\section{Discussion \& Conclusion}\label{Discussion}
\textbf{Discussion.} This paper analyzes and formulates the extended MSME AD task for the first time and outlines a possible concept with dedicated potential solutions along related open questions. Due to the novel scope of the extended problem, no baseline is available and therefore, a quantification of efficiency gains is not possible. In addition, the fragmentation or cohesion of partnerships and collaborations across vendors, countries, and other stakeholders has a significant impact. The degree of collaboration and partnerships is unpredictable. However, it is unlikely that an all-encompassing collaboration can be achieved. Yet, even in the worst case, a single vendor-specific concept implementation, the gains are foreseeably large. 

The development of reward models and the social discussion of critical decisions, on the other hand, can be consolidated in country-specific settings, decoupled from vendors. In this way, complex socio-political matters can be addressed in the necessary breadth and integrated natively into the techno-specific development. In addition to this native integration of extended externalized requirements, system-specific transfer learning enables a native design choice validation approach, which also minimizes risks and streamlines overall iterative refinement and alignment.

However, there are still numerous open questions that remain critical. Two major challenges are the exploration of the boundary between generalization and adaptation and the regional breakdown. In this context, it is particularly important to investigate how similarity can be assessed and, ultimately, what level of granularity is required for country-specific reward models. Furthermore, cost factors are currently underrepresented. However, these cannot yet be analyzed in detail until the aforementioned aspects are better understood. Accordingly, the development of country-specific reward models and their separation must first be examined, e.g., in the context of planning, and then gradually extended across the entire stack.

\textbf{Conclusion.} Overall, this paper addresses the challenge of scaling AD while considering the recent AD stack evolution along process paradigm shifts and data-driven implicit concerns, such as dataset incompleteness, the open long-tail distribution of the real world, and dynamic shifts across space and time. The proposed concept tackles the broader MSME task, ensures continuous improvement, enables cross-border sharing of catastrophic failure data, and enhances traceability, public accessibility, and clarity. It also raises open research questions, including automated scenario extraction, database integration, and country-specific reward models. By integrating technical, legal, cultural, and ethical aspects, the approach supports efficient scalability while reconsidering data-driven AD at scale. In doing so, the paper contributes to the advancement of AD and highlights key challenges alongside potential solutions and research directions.

\bibliographystyle{IEEEtran}
\bibliography{literature}

\end{document}